\documentclass[prx,twocolumn]{revtex4-1}

\usepackage{graphicx,hyperref,bm,bbm,amssymb,amsfonts,amsmath}

\newcommand{\sz}[1]{\sigma^{\rm z}_{#1}}
\def\ii{{\rm i}}
\newcommand{\ket}[1]{|{#1}\rangle}
\newcommand{\bra}[1]{\langle {#1}|}

\newcommand{\bracket}[3]{\langle{#1}|{#2}|{#3}\rangle}
\newcommand{\tr}{\mathrm{tr}}
\newcommand{\ave}[1]{\langle{#1}\rangle}
\newcommand{\e}[1]{{\rm e}^{#1}}

\date{\today}

\begin{document}

\title{Entanglement in a dephasing model and many-body localization}
\author{Marko \v Znidari\v c}
\affiliation{Physics Department, Faculty of Mathematics and Physics, University of Ljubljana, 1000 Ljubljana, Slovenia}

\begin{abstract}
We study entanglement dynamics in a diagonal dephasing model in which the strength of interaction decays exponentially with distance -- the so-called l-bit model of many-body localization. We calculate the exact expression for entanglement growth with time, finding in addition to a logarithmic growth, a sublogarithmic correction. Provided the l-bit picture correctly describes the many-body localized phase this implies that the entanglement in such systems does not grow (just) as a logarithm of time, as believed so far.
\end{abstract}

\maketitle

\section{Introduction}
Localization is a phenomenon that, due to its peculiar properties, is of interest in different fields of physics. As its name already implies, one of the characteristic properties is a lack of transport and as such it was first considered within solid-state questions of transport. Somewhat surprisingly Anderson found~\cite{Anderson58} that in one-dimension and for noninteracting particles an infinitesimal disorder causes an abrupt change of all eigenstates from extended to localized. Being an interference phenomenon one could argue that any interaction between particles will wash out precise phase relations and thereby destroy localization. That this needs not be so was shown using diagramatics in Ref.~\cite{BAA}, see also Ref.~\cite{Gornyi}. A couple of numerical works followed~\cite{Vadim07,ZPP08}, realizing that such many-body localized (MBL) systems display many interesting properties~\cite{Monthus10,Pal10}. This eventually led to a flurry of activity in recent years, see Ref.~\cite{rew} for a review.

One of the characteristic features of MBL systems is its logarithmic in time growth of entanglement entropy~\cite{ZPP08} (in a finite system the entropy growth will eventually stop at a volume-law saturation value~\cite{Bardarson12}). Although conserved quantities like energy or particles are not transported in MBL systems, quantum information/correlations do spread, which is in contrast to a single-particle (i.e., Anderson) localization where the entropy does not grow. Logarithmic growth has been explained early on as being caused by a dephasing due to exponentially decaying effective interaction~\cite{Vosk13,Serbyn13}, see also Ref.~\cite{Kim14}, with a prefactor that is equal to the localization length~\cite{Serbyn13,Serbyn14,Vadim14,Kim14,Nanduri14}. This picture has been furthermore elaborated by a so-called l-bit picture~\cite{Serbyn14,Vadim14} (also called local integrals of motion (LIOM) picture) that nowadays constitutes what is believed to be the fullest description of MBL. It relies on an existence of (quasi) local integrals of motion such that a quasi-local unitary transformation can change an MBL Hamiltonian from its physical basis (real space) to a logical l-bit basis where $H$ is diagonal, and can be written as,
\begin{equation}
H=\sum_k J^{(1)}_k \sz{k}+\sum_{k<l} J^{(2)}_{k,l} \sz{k}\sz{l}+\sum_{k<l<m} J^{(3)}_{k,l,m} \sz{k}\sz{l}\sz{m}+\cdots,
  \label{eq:H}
\end{equation}
with $\sz{k}$ being the Pauli matrix at the $k$-th l-bit site. The coupling constants $J^{(r)}$ are inhomogeneous and implicitly depend on the disorder in the original model. Because it is diagonal the l-bit Hamiltonian (\ref{eq:H}) manifestly displays the emergent effective integrability of MBL systems~\cite{Chandran,Ros}, another reason for their high interest. While for most systems that are believed to be MBL the existence of the l-bit description is in principle a conjecture, its construction is implicit in a proof of MBL for a particular system~\cite{Imbrie16}. It is also able to describe many phenomenological properties of MBL systems~\cite{ImbrieRew} and is as such widely believed to be the correct description of MBL.

Still, considering a vast amount of predominantly numerical works (see though e.g. Refs.~\cite{Imbrie16,Eisert15,Eisert15b} for exact results) discussing MBL in general, as well as entanglement specifically, e.g. being at a core of a definition of MBL~\cite{Bauer13}, its behavior in the MBL phase or close to the transition~\cite{Vosk13,Vosk15,Goold15,Serbyn15,Arul16,Luitz16,Iemini16,Pixley17,Tomasi17}, in long-range~\cite{Pino14,Roy17} and time-dependent~\cite{Potter18} systems, or for bond disorder~\cite{Sirker16}, it would be extremely useful to have analytical results for MBL systems, or for its conjectured canonical l-bit form (\ref{eq:H}). Our goal is to provide such a result for entanglement growth. 

We first in Sec.II. discuss the saturation value of entanglement at long times using only ergodicity of eigenvalues of $H$, without invoking any disorder. This shows that the initial state that gives the largest saturation value, and therefore will exhibit the longest logarithmic growth before eventually saturating, is an initial state with zero expectation value of magnetization in the $z$ direction. In subsequent sections we then calculate the entanglement growth for such optimal initial states and for a Gaussian distribution of coupling constants whose size decays exponentially with the distance -- a usual assumption for the l-bit model. In Sec.~III. we solve the 2-body l-bit model (the one having only 1- and 2-body interactions in (\ref{eq:H})) and calculate the explicit time dependence of the Reny-2 entropy (purity). This section gives our main result -- the entropy growth is not just logarithmic in time but instead has a sub-leading correction. Dependence of all the constants on localization length is explicit. While the exact result is for a particular initial state, we argue and numerically demonstrate that our result is robust with respect to different generic initial states, different distribution of couplings, and other Reny or von Neumann entropies. In Sec.~IV. we then discuss the 3-body model, again getting a similar result as for the 2-body model, indicating that the form of the sub-leading correction does not depend on the order of interactions. In Sec.~V. we explain how the result for the l-bit model directly translates to a putative MBL system in the physical basis. Finally, in Sec.~VI. we numerically study the leading order time evolution of individual eigenvalues, fully specifying the entanglement content of an evolved state, finding a nonmonotonic convergence to the asymptotic random-state eigenvalues.

\section{Diagonal dephasing model and the saturation entanglement}

The local integrals of motion model, in short the l-bit model, is given by a diagonal Hamiltonian in Eq.(\ref{eq:H}) where one assumes that $J^{(r)}_{k,l,\ldots}$ decay exponentially with increasing maximal distance between their site indices $k,l,\ldots$, as well as with the increasing order $r$~\cite{Serbyn14,Vadim14}. System length is $L$, Hilbert space size $N=2^L$, while $k,l,m$ will be site indices. We would like to calculate time evolution of entanglement described by such $H$. For now we leave the precise values of $J^{(r)}_{k,l,\ldots}$ unspecified as they are not needed for the calculation of the asymptotic saturation value of entanglement. 

At first sight the problem of time evolution might appear trivial -- after all $H$ is diagonal with eigenstates being just the basis (computational) states. While the evolution is indeed trivial (no evolution) if one starts with a single basis state (in the l-bit basis), situation can be rather complex if one starts with a {\em superposition} of basis states (even if they represent a product initial state). Complexity in quantum mechanics can come not just from the dynamics but also from the complexity of an initial state. One can in fact ask what is the complexity of simulating evolution by diagonal matrices, in other words of the dynamics governed only by phases (commuting operators). The answer is not known, though it is believed~\cite{Bremner11} that such circuits in general can not be simulated efficiently on a classical computer. MBL systems are in this sense not generic as their entanglement grows logarithmically with time~\cite{foot1}, and thus the simulation complexity polynomially with time (low entanglement is a sufficient, but not necessary, condition for an efficient simulatability).

Starting from a pure state $\ket{\psi(0)}=\sum_{p=1}^{N} c_p \ket{p}$ we would like to calculate the entanglement in a state after time $t$, $\ket{\psi(t)}={\rm e}^{-\ii H t}\ket{\psi(0)}$, where we set $\hbar=1$. For pure states the bipartite entanglement is fully specified by the spectrum of the reduced density matrix $\rho_{\rm A}(t)=\tr_{\rm B} \ket{\psi(t)}\bra{\psi(t)}$. A convenient measure is purity, $I(t)=\tr \rho_{\rm A}^2$, and closely related Reny-2 entropy $S_2(t):=-\log_2 I(t)$. In all our calculations we shall calculate the average $I(t)$ and then take its logarithm to get $S_2(t)$, arguing that $S_2(t)$ for large times and in the thermodynamic limit (TDL) behaves essentially the same as the von Neumann entropy, and furthermore, due to self-averaging taking the logarithm of the average is essentially the same as taking the average of the logarithm. For a particular model studied this is demonstrated numerically in the Appendix.

While the eigenstates of $H$ are simple, the eigenenergies are combinations of various $J^{(r)}_{j,k,\ldots}$ depending on the orientations of individual spins. Let us denote those eigenstates by $E_{\bf{j}\bm{\alpha}}$, where we shall use a double (multi)index labeling bipartite eigenstates $\ket{\bf{j}}_{\rm A}\otimes \ket{\bm{\alpha}}_{\rm B}$, that is ${\bf{j}}\equiv(j_1,j_2\cdots,j_{L_{\rm A}})$ and $\bm{\alpha}\equiv(\alpha_{L_{\rm A}+1},\alpha_{L_{\rm A}+2},\ldots,\alpha_{L_{\rm A}+L_{\rm B}})$ with a binary $j_k \in \{+1,-1\}$ and $\alpha_l \in \{+1,-1\}$ labeling the state of the l-bit and we use a bipartition into $L_{\rm A}+L_{\rm B}=L$ sites. From now on we use roman $i,j$ and Greek $\alpha,\beta$ as eigenstate (multi)indices on the respective subspaces, dropping the vectorial notation on them. Calculating the purity one gets
\begin{equation}
  I(t)=\sum_{i,j,\alpha,\beta} c_{i\alpha} c_{j\alpha}^* c_{j\beta} c_{i\beta}^* \e{-\ii (E_{i\alpha}-E_{j\alpha}+E_{j\beta}-E_{i\beta})t}.
\label{eq:Isum}
  \end{equation}

\subsection{Saturation value}

Let us first calculate the asymptotic saturation value of $I(t \to \infty)$. Assuming the eigenenergies are ergodic (which is for instance the case for our 2-body model studied later), performing an infinite time averaging one has $\overline{\exp{(-\ii (E_{i\alpha}-E_{j\alpha}+E_{j\beta}-E_{i\beta})t)}}=\delta_{ij}+\delta_{\alpha \beta}-\delta_{ij}\delta_{\alpha \beta}$, resulting in $\overline{I(t)}=\sum_{i,\alpha,\beta} |c_{i\alpha}|^2|c_{i\beta}|^2+\sum_{i,j,\alpha} |c_{i\alpha}|^2|c_{j\alpha}|^2-\sum_{i,\alpha}|c_{i\alpha}|^4$ (similar calculations have been used many times~\cite{Nakata12,Karol}). Taking an initial product state $\ket{\psi(0)}=(\cos{\frac{\varphi}{2}}\ket{0}+\sin{\frac{\varphi}{2}}\ket{1})^{\otimes L}$ gives a saturation value of purity for a bipartition into $L_{\rm A}=L-L_{\rm B}$ consecutive sites $\overline{I(t)}=q^{L_{\rm A}}+q^{L_{\rm B}}-q^L$, where $q:=\frac{3+\cos{2\varphi}}{4}=\frac{1+z^2}{2}$, and $z:=\bracket{\psi(0)}{\sz{k}}{\psi(0)}$. Focusing on an equal bipartition, $L_{\rm A}=\frac{L}{2}$, the expression for the saturation value of $S_2$ simplifies to
\begin{equation}
\overline{S_2(t)}=c \frac{L}{2}-1,\qquad c:=\log_2\frac{2}{1+z^2}.
\end{equation} 
The saturation value of the entropy therefore always satisfies a volume law, with a prefactor $c$ being the larger the smaller is the value of the initial $z$ (formula for $c(z)$ explains numerical observation in Ref.\cite{Nanduri14}). From an experimental (real or numerical) point of view it is therefore best to choose the initial state with $z=0$ -- this will result in the largest saturation value and therefore the largest range of values where the entropy growth can be observed. With that aim we shall focus on the initial state with $\varphi=\frac{\pi}{2}$ ($z=0$), i.e., $c_{j\alpha}=1/\sqrt{N}$ (we shall also show that random product initial states give the same behavior).

\section{Random 2-body model and time evolution}

Let us now return to our main focus -- the time dependence of $I(t)$. The simplest case of the l-bit dephasing model (\ref{eq:H}) is a so-called 2-body model for which
\begin{equation}
J^{(r)}_{k,l,\ldots}\equiv 0\qquad \mbox{for all $r>2$},
\end{equation}
while $J^{(1)}_k$ and $J^{(2)}_{k,l}$ are nonzero. Their precise form will be specified later. Such a model is simpler for analytical treatment while, as we will argue, still retains all the features of the full model (\ref{eq:H}). We want to describe time evolution with such $H$.

Taking the optimal initial state with $c_{j\alpha}=1/\sqrt{N}$ we need in Eq.~(\ref{eq:Isum}) only the differences of eigenenergies $E_{i\alpha}-E_{j\alpha}+E_{j\beta}-E_{i\beta}$. For the 2-body model those simplify, such that the only terms remaining are the 2-body $J^{(2)}_{k,l}$ that couple one site from the subsystem ${\rm A}$ and one from ${\rm B}$. The expression for purity that one gets is
\begin{equation}
  I(t)=\frac{1}{N^2}\sum_{{\bf i},{\bf j} \in {\rm A},\bm{\alpha},\bm{\beta} \in {\rm B}} \e{-\ii t (\mathbf{i}-\mathbf{j})\cdot J^{(2)}_{\rm AB}\cdot (\bm{\alpha} - \bm{\beta})},
  \label{eq:IAB}
\end{equation}
where, to avoid confusion, we temporarily re-introduce boldface multiindices labeling the basis states, and $J^{(2)}_{AB}$ is a $L_{\rm A}\times L_{\rm B}$ matrix containing all the 2-body couplings between subsystems ${\rm A}$ and ${\rm B}$. Purity (\ref{eq:IAB}) is still a sum over exponentially many terms ($N^2$ in number), but the argument of the exponential function involves only $L_{\rm A}L_{\rm B}$ terms. For instance, for $L_{\rm A}=L_{\rm B}=2$ and consecutive sites, the argument of the exponential function in Eq.(\ref{eq:IAB}) has $4$ terms and is proportional to
\begin{equation}
(i_1-j_1,i_2-j_2)\begin{pmatrix} J^{(2)}_{1,3} & J^{(2)}_{1,4} \\
J^{(2)}_{2,3} & J^{(2)}_{2,4}
\end{pmatrix}
\begin{pmatrix}
\alpha_3-\beta_3\\
\alpha_4-\beta_4
\end{pmatrix}.
\label{eq:ex22}
\end{equation}
Eq.~(\ref{eq:IAB}) is the central formula that we build upon. 

Let us now introduce a {\em random} 2-body model in which the distribution of $J^{(2)}_{k,l}$ is Gaussian and independent for each pair of sites $k,l$. Specifically, the distribution is $p(x=J^{(2)}_{k,k+r})\sim \e{-x^2/2J^2 W_r^2}$, that is with zero mean and the variance
\begin{equation}
\ave{(J^{(2)}_{k,k+r})^2}=J^2 W_{r}^2:=J^2 {\rm e}^{-2(r-1)/\xi}.
\label{eq:J2}
\end{equation} 
The size of the coupling decays exponentially with distance, the decay length being $\xi$, while $J$ sets the energy (time) scale. We are predominantly interested in the long-time behavior when the entanglement is large (i.e., volume law that is a (small) fraction of $\sim L$) and the states involved are thus generic. Therefore one expects that relative sample-to-sample fluctuations decrease with time and can be neglected. Averaging purity (\ref{eq:IAB}) over Gaussian distribution of couplings one gets the average purity,
\begin{equation}
  \ave{I(t)}= \frac{1}{N}\sum_{i,\alpha} \prod_{k \in A,l\in B} \e{-8J^2t^2 W^2_{l-k} \delta_{i_k,1} \delta_{\alpha_l,1} }.
  \label{eq:I}
  \end{equation}
For a non-Gaussian distribution of $J^{(2)}$ one would have in Eq.(\ref{eq:I}) instead a product of Fourier transformations of the distribution $p(x)$. Essentials would be the same (see Appendix~\ref{app1}). While the expression (\ref{eq:I}) is simpler than (\ref{eq:IAB}), being a sum of $N$ instead of $N^2$ terms, it is still combinatorially complex. In the sum over $2^L$ bit strings $\ket{i,\alpha}$ each pair of bits $k,l$ (from parts A and B, respectively) contributes a term $\sim W^2_{l-k}$ in the exponential argument if the $k$-th and $l$-th bits are $+1$. While many of $N$ bit strings result in the same argument of the exponential, there are still of order $\approx 0.4 N$ different terms and the expression therefore can not be much more simplified provided one wants to retain its exactness.

To give an idea of the form that $\ave{I(t)}$ takes we write the exact expression for $L=4$ and $L_{\rm A}=2$ that is obtained by averaging Eq.~(\ref{eq:IAB}) with (\ref{eq:ex22}) using (\ref{eq:J2}),
\begin{widetext}
\begin{equation}
  \ave{I(t)}=\frac{1}{16}\left[ 7+\left(\e{-\tau^2 W_1^2}+2\e{-\tau^2 (W_1^2+W_2^2)}+\e{-\tau^2(W_1^2+2W_2^2+W_3^2)}\right)+\left(2\e{-\tau^2 W_2^2}+2\e{-\tau^2(W_2^2+W_3^2)}\right)+\left(\e{-\tau^2 W_3^2}\right) \right],
  \label{eq:L4}
\end{equation}
\end{widetext}
where $\tau^2:=8J^2t^2$. Round brackets group terms whose leading argument is the same $W_r^2$, $r=0,1,2,3$. We have generated such exact expressions for an equal bipartitions for up-to $L=34$ spins (where $\ave{I(t)}$ is a sum of the order of $\sim 10^{10}$ exponential functions with different arguments). While they are obviously too long to be written out, their general form is
\begin{equation}
\ave{I(t)}=\!\! \frac{1}{N}\!\!\left[\!(2^{1+\frac{L}{2}}-1)+\sum_{r=1}^{L-1} \sum_{m} d^{(r)}_m \, \e{-\tau^2\sum_{p=r}^{L-1} c^{(r)}_{m,p} W_p^2} \right].
\label{eq:exact}
\end{equation}
The first constant term is just the saturation value giving the already mentioned $-\log_2 \overline{\ave{I(t)}} = \frac{L}{2}-1-\log_2{(1-2^{-L/2-1})}$. One can show that the leading coefficient is $c^{(r)}_{m,r}=1$ for all $r$ (also, trivially, $c^{(r)}_{m,p}\le p$, because there can be at most $p$ links of length $p$ across a given cut).

In the next two subsections we shall discuss the asymptotic closed-form expressions for the purity decay obtained by replacing sums with integrals. We shall first discuss the case of small localization lengths where physics as well as mathematical derivations are rather transparent. Then we are going to derive expressions that hold also for larger localizations lengths (as well as for small), leading to essentially the same expression as in the simpler case of small localization lengths.

\subsection{Small localization length}

Let us first discuss the case of small $\xi$ where analytic treatment is the simplest. Because $W_r^2$ decrease exponentially with $r$ like $W_r^2 = \e{-2(r-1)/\xi}$ we can in each of $N$ terms in Eq.(\ref{eq:I}) retain in the argument only terms with the smallest $r$ in $W_{r=l-k}^2$ (we can do that because $c_{m,p}^{(r)}$ grow at most linearly with $p$), see also the explicit example in Eq.~(\ref{eq:L4}). Small localization length therefore means $\xi \ll 1$; after deriving general expression in the next subsection we will see though that in practice having $\xi \lesssim 1$ is enough.

\begin{figure}[t!]
\centerline{\includegraphics[width=3.0in]{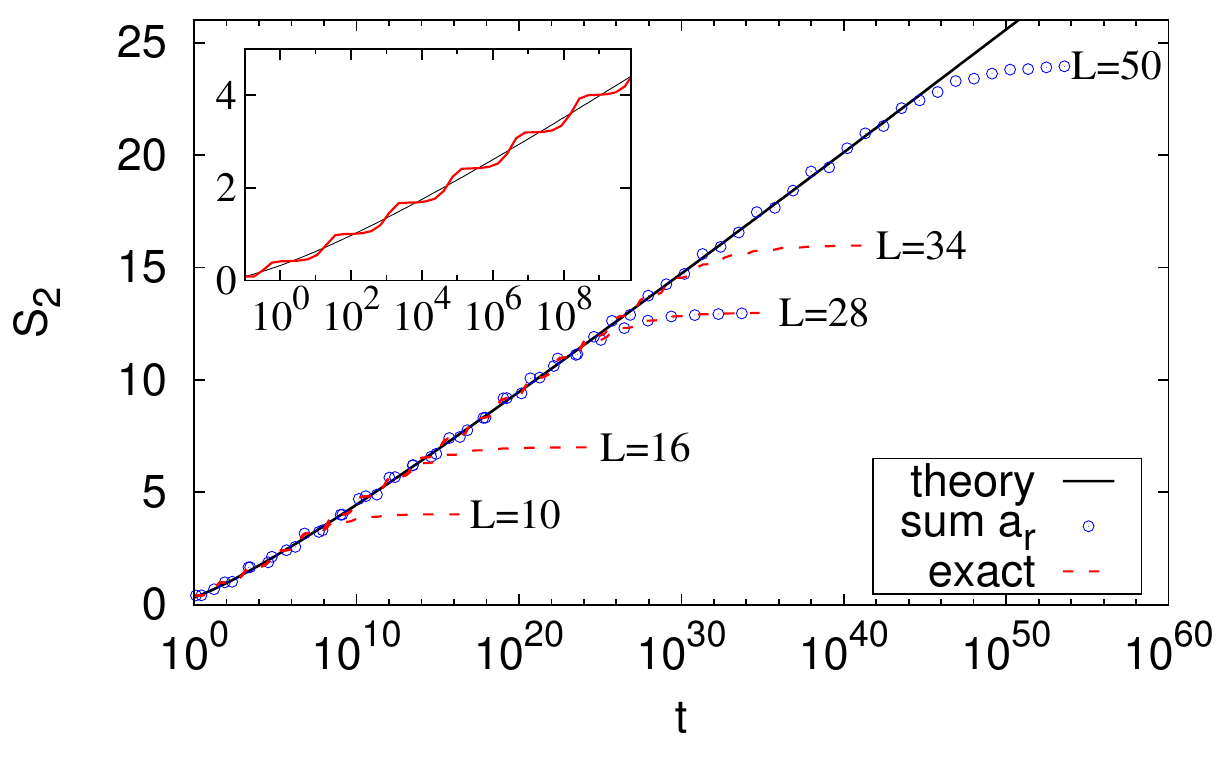}}
\caption{(Color online) Entanglement growth in a 2-body random dephasing model for small localization length $\xi=0.25$, $S_2=-\log_2 I(t)$. Full curve is Eq.~(\ref{eq:Iplat}), circles the sum (\ref{eq:ar}), and red curves the exact Eq.~(\ref{eq:I}). The inset shows a small time zoom-in, showing plateaus.}
\label{fig:small}
\end{figure}

The number of leading terms, denoted by $a_r:=\sum_m d_m^{(r)}$, can be calculated exactly. Looking at (\ref{eq:I}) we have to consider contributions from $2^L$ possible bit strings of length $L$. If we are interested in terms that have a minimal distance $r$ (i.e., leading order $W_r^2$) it is enough to consider $r+1$ bits around the bipartite cut. Bits that are $-1$ (remembering that we use a conventon where the multiindex bits, e.g. $i_k$, take values $+1$ and $-1$) prevent the corresponding bond term to appear in (\ref{eq:I}). Therefore, we just have to count the number of such bit strings that have for each pair of bits (one from A, one from B) at distance smaller than $r$ at least one bit set to $-1$. This is obtained by a series of $r-1$ bits set to $-1$ followed by $1$ at each end, e.g., for $r=4$ one has $\ldots 1(-1)(-1)(-1)1\ldots$, and because we can put a cut at $r$ different positions between these $r+1$ highlighted bits, we immediately get $a_r=r 2^{L-r-1}$ (all $L-r-1$ nonhighlighted bits can have arbitrary values because they contribute to subleading terms). This holds as long as we are away from the boundaries, that is for $r \le L/2$. Taking into account also the boundaries one arrives at $a_r=r2^{L-r-1}$ if $r \le L/2$, while $a_r=(r-2(r-\frac{L}{2}))2^{L-r-1}$ otherwise. As an example, for $L=4$ we have $a_1=4$, $a_2=4$ and $a_3=1$ (compare with the explicit (\ref{eq:L4})). For small $\xi$ we can therefore write
\begin{equation}
  \ave{I(t)} \approx \frac{2}{2^{L/2}} +\sum_{r=1}^{L-1} \frac{a_r}{N} \e{-\tau^2 W_r^2}.
  \label{eq:ar}
\end{equation}
For finite $L$ the terms with $a_{r \le L/2}$ will contribute in the first half (in logarithmic scale) of the decay to the asymptotic saturation, while smaller terms $a_{r>L/2}$ kick in only in the second half. As we are interested in the behavior in the thermodynamic limit we can safely make the limit $L \to \infty$ for any fixed $t$, obtaining purity decay in the thermodynamic limit and small $\xi$,
\begin{equation}
\ave{I(\tau)}\approx \sum_{r=1}^\infty \frac{r}{2^{r+1}} \e{-\tau^2 W_r^2},\quad \tau:=Jt\sqrt{8},
\label{eq:Iar}
\end{equation}
where $\tau$ is a convenient time parameter, and we recall $W_r=\e{-(r-1)/\xi}$. For small $\xi$ the values of $W_r$ greatly differ for different $r$ and so it follows that $\ave{I(t)}$ decays (and $S_2$ grows) in a series of steps (see Fig.\ref{fig:small}) connecting plateaus in purity. The value of the $r$-th plateau is at $I_r=1-\sum_{m=1}^r \frac{m}{2^{m+1}}=\frac{r+2}{2^{r+1}}$ (e.g., $I_r=\frac{3}{4},\frac{4}{8},\frac{5}{16},\ldots$ for $r=1,2,3,\ldots$). The sum in (\ref{eq:Iar}) still obscures the time dependence of $I(t)$. To get a better understanding we study at what times different plateaus are reached. The transition from one to the next plateau happens when the argument $\tau_r^2 W_r^2 \approx 1$, that is at a time satisfying $r-1 \approx \xi \ln{\tau_r}$, at which the value of $-\log_2\ave{I(t)}$ is around $\frac{1}{2}(-\log_2 I_{r-1}-\log_2 I_r) \approx r+\frac{1}{2}-\log_2(r+\frac{3}{2})$. Putting the two together results in (expression is expected to be valid for large $r$, i.e., large $\xi \ln\tau$)
\begin{equation}
S_2(\tau) \approx \xi \ln\tau+A(\xi)-\log_2{\left[B(\xi)+\xi \ln \tau\,\right]},
\label{eq:Iplat}
\end{equation}
with $A(\xi)=\frac{3}{2}, B(\xi)=\frac{5}{2}$. We keep an explicit dependence of $A,B$ on $\xi$ because, as we shall show, the same form with $\xi$-dependent constants is obtained also at larger $\xi$. Eq.(\ref{eq:Iplat}), showing that entanglement does not grow as a simple logarithm of time, is our main result. While the leading logarithmic dependence has been observed and heuristically explained before, we also get a new negative logarithmic correction $\log_2{[B(\xi)+\xi \ln \tau]}$.

Taking instead of the mean plateau either $I_{r-1}$ or $I_r$ we get a lower/upper bound on $-\log_2\ave{I(t)}$ for which $A(\xi)=1\,(2)$ and $B(\xi)=2\,(3)$ for the lower (upper) bound. From the derivation it is clear from where does the subleading log-log correction come: it is due to the numerator $r$ in Eq.(\ref{eq:Iar}) which is in turn related to the linear growth of the number of different possible couplings of length $r$ crossing the cut. For small $r$ there are simply fewer connections, and therefore at short times when small $r$ matter, entanglement growth is slightly slower than at larger times. It is therefore a robust feature independent of a particular 2-body model (the denominator $2^{r+1}$ on the other hand comes from the Hilbert space size of all states connected with bonds of length $\le r$).

\begin{figure}[th!]
\centerline{\includegraphics[width=2.8in]{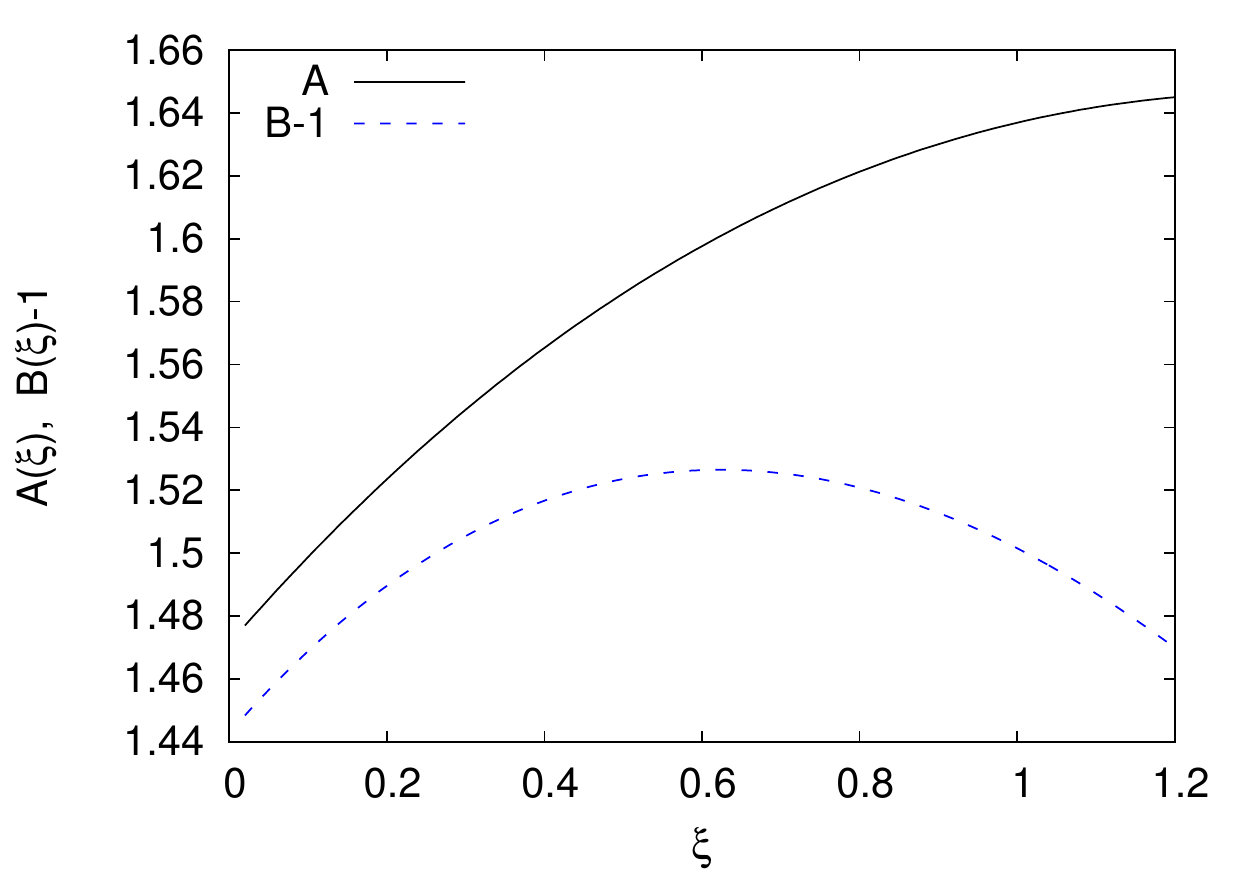}}
\caption{(Color online) Dependence of $A(\xi)$ and $B(\xi)$ in Eq.(\ref{eq:Iplat}) for small localization length $\xi$ as given by Eq.~(\ref{eq:ABsmall}).}
\label{fig:AB1}
\end{figure}

One can also give an alternative analytic derivation of the logarithmic correction. Replacing the sum in Eq.(\ref{eq:Iar}) with an integral over $r$, in turn changing the variable $r$ to $y:=\tau^2 \e{-2(r-1)/\xi}$, one gets
\begin{equation}
\ave{I(\tau)} \approx \frac{\xi}{8} 2^{-\xi \ln{\tau}}\int_0^{\tau^2} \frac{1+\xi \ln\tau-\frac{\xi}{2}\ln{y}}{y^{1-\frac{\xi}{2}\ln{2}}} \e{-y}{\rm d}y.
\label{eq:Iint}
\end{equation} 
This integral is very handy for deriving the asymptotic expansion valid for $\xi \ln{\tau} \gg 1$. Namely, at the upper limit of integration the integrated function is exponentially small and we can safely extend the upper limit of integration to infinity. The resulting integral is elementary end equal to $\Gamma(\frac{\xi}{2}\ln{2})\left(1+\xi \ln\tau-\frac{\xi}{2} \Psi(\frac{\xi}{2}\ln{2})\right)$, where $\Gamma(z)$ is the Gamma function and $\Psi(z):=\Gamma'(z)/\Gamma(z)$ the Digamma function. Taking a negative logarithm of purity to get $S_2$ the expression has the same form as in Eq.(\ref{eq:Iplat}), with
\begin{equation}
A(\xi)=-\log_2{\left[\frac{\xi}{8}\Gamma\left(\frac{\xi}{2}\ln{2}\right)\right]},\, B(\xi)=1-\frac{\xi}{2} \Psi\left(\frac{\xi}{2}\ln{2}\right).
\label{eq:ABsmall}
\end{equation}
Compared to the values of $A$ and $B$ obtained from the simplistic plateau analysis we here also have $\xi$-dependent corrections ($B$ can also be expressed as $B(\xi)=\xi \frac{{\rm d}A(\xi)}{{\rm d}\xi}+1+1/\ln{2}$), with the limiting values $A(\xi \to 0)\approx 1.47$ and $B(\xi \to 0)\approx 2.44$, see Fig.~\ref{fig:AB1}. In Fig.~\ref{fig:small} we show comparison of the exact $S_2(t)$ (\ref{eq:I}), the approximate sum (\ref{eq:ar}), and analytic result Eq.(\ref{eq:Iplat}) with $A(0.25)\approx 1.53$ and $B(0.25)\approx 2.50$ obtained from Eq.(\ref{eq:ABsmall}). Excellent agreement is observed.

\subsection{General localization length}

\begin{figure}[t!]
\centerline{\includegraphics[width=2.8in]{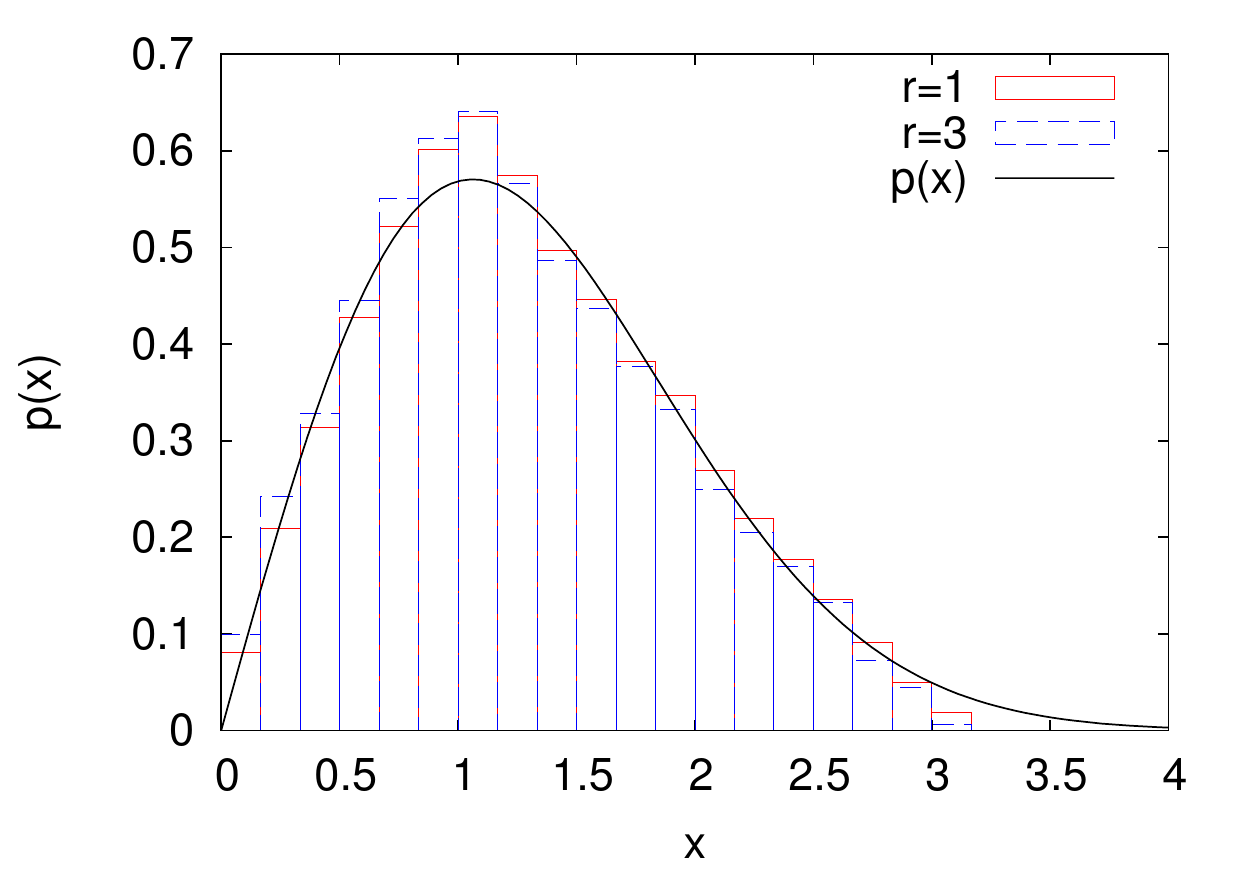}}
\caption{(Color online) Distribution of $x$ (see text for definition) for $L=14$, $\xi=3$, and $r=1$ (distribution is over all $a_1=4096$ terms with $r=1$), as well as $r=3$ ($a_3=3072$). Full curve is heuristic $p(x)$ that we use (see text), with $b\approx 2.26$ determined from theoretical value of $\bar{x}$.}
\label{fig:podx}
\end{figure}

\begin{figure}[th!]
\centerline{\includegraphics[width=2.8in]{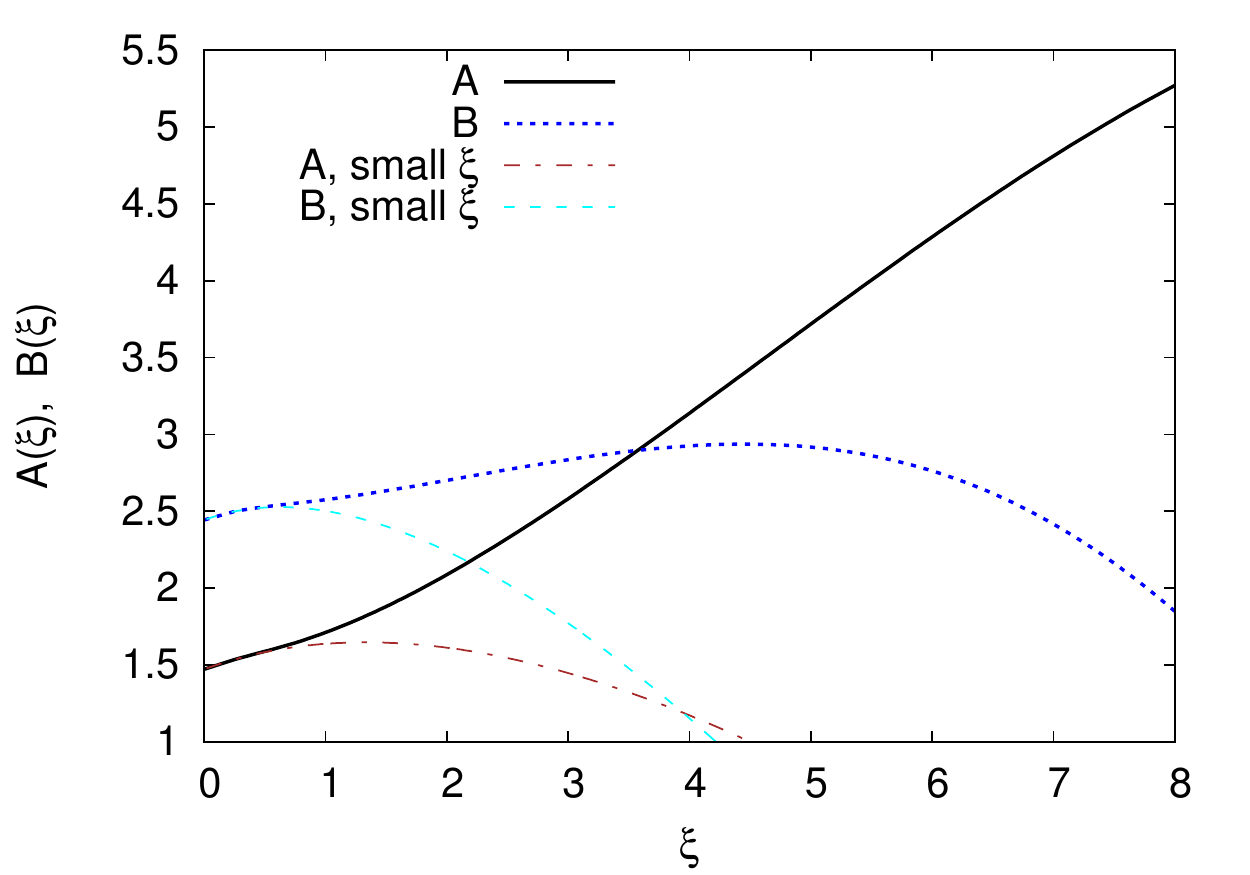}}
\caption{(Color online) Dependence of $A(\xi)$ and $B(\xi)$ as given by Eq.~(\ref{eq:ABpor}) which holds for generic $\xi$. For comparison we also show (dashed and chain curves) the values obtained for small $\xi$ (\ref{eq:ABsmall}) shown in Fig.~\ref{fig:AB1}.}
\label{fig:AB2}
\end{figure}

\begin{figure}[t!]
\centerline{\includegraphics[width=2.7in]{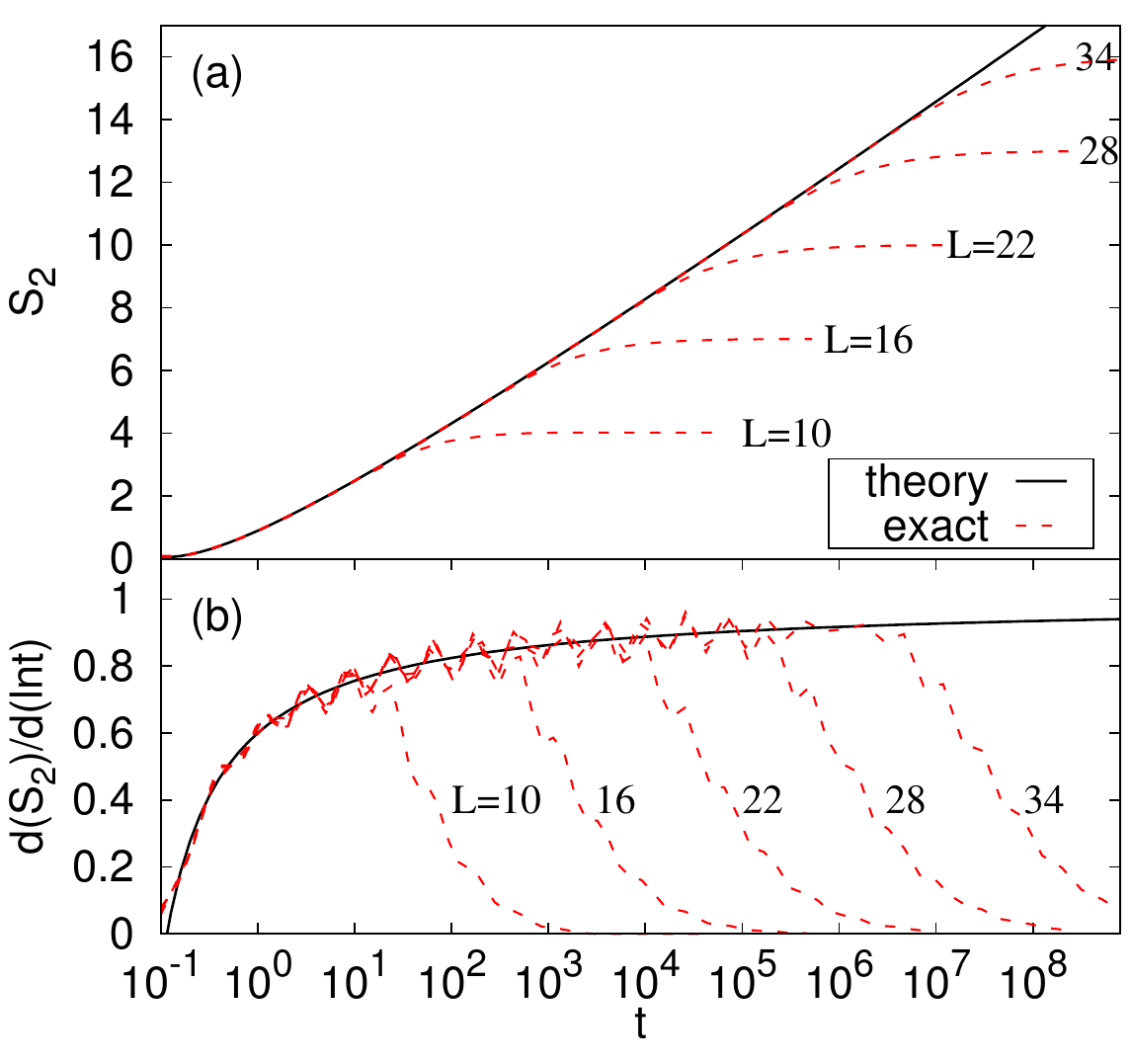}}
\centerline{\includegraphics[width=2.7in]{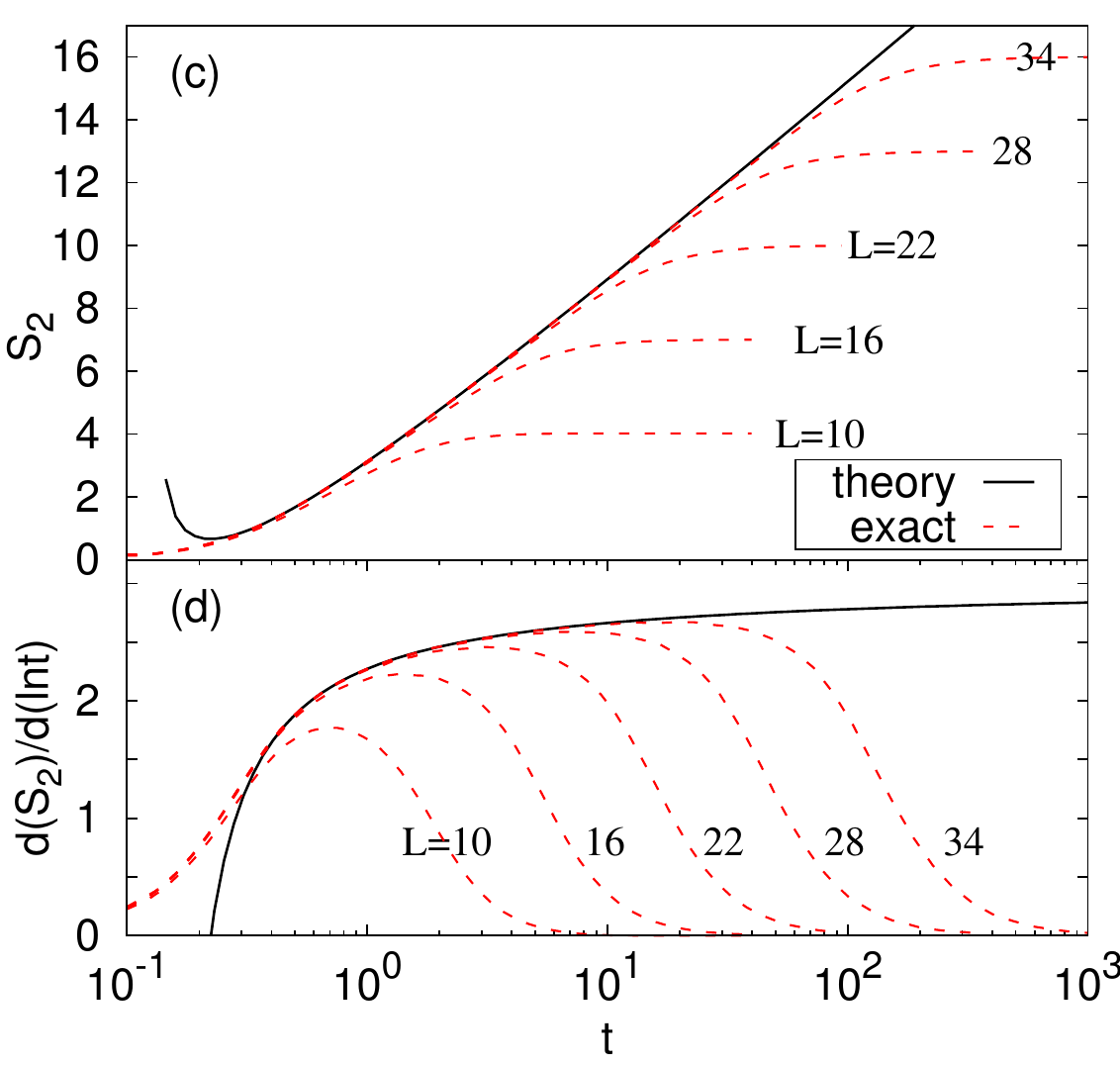}}
\caption{(Color online) Entanglement growth in a 2-body random dephasing model for $\xi=1$ in (a) and (b), and $\xi=3$ in (c) and (d). (b) and (d) show logarithmic derivatives, clearly indicating the presence of a logarithmic correction (\ref{eq:ABpor}). Theory here represents (\ref{eq:ABpor}) with $A(1)\approx 1.71$, $B(1)\approx 2.57$, and $A(3)\approx 2.58$, $B(3)\approx 2.83$.}
\label{fig:large}
\end{figure}

\begin{figure}[thb!]
\centerline{\includegraphics[width=3in]{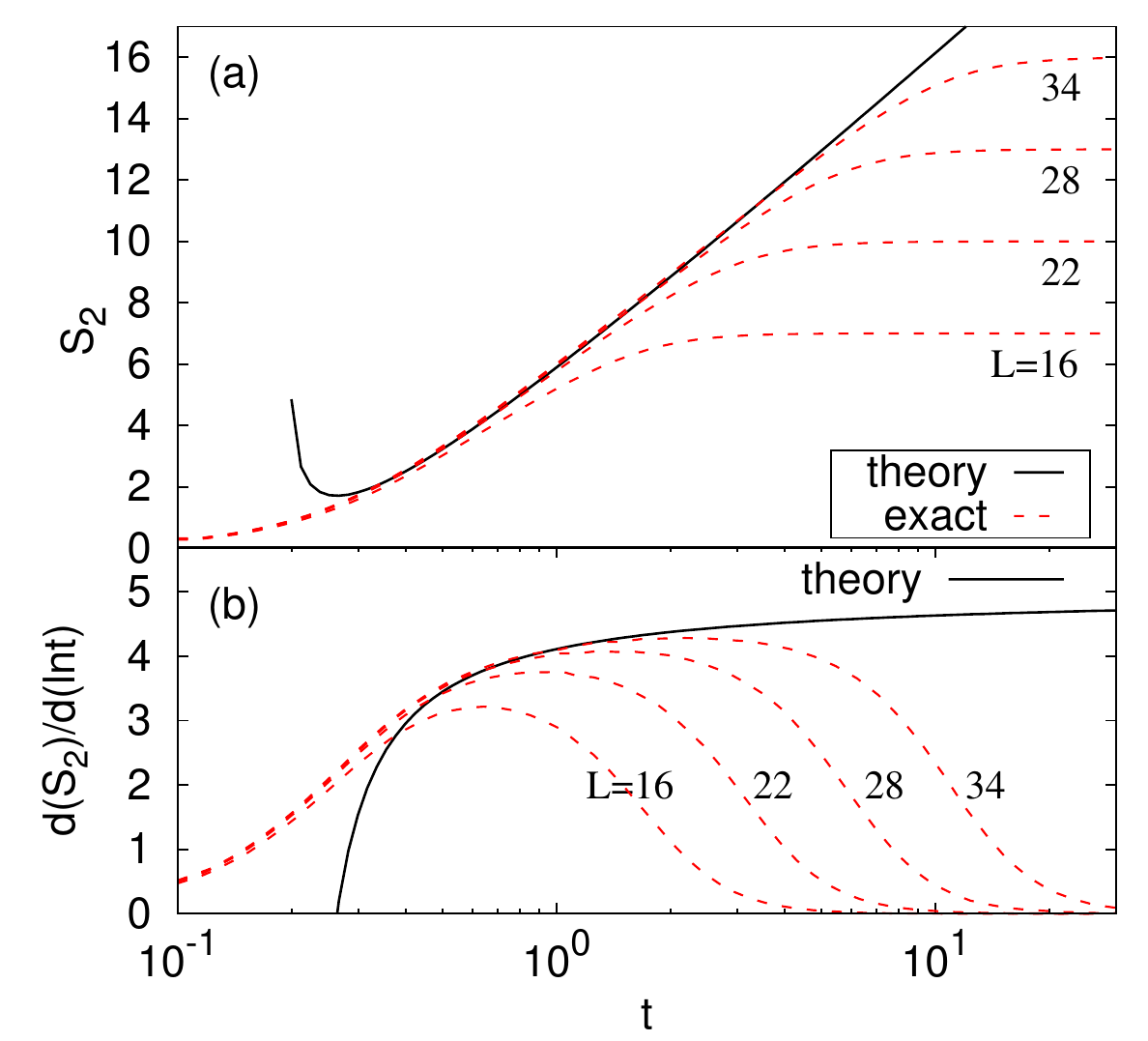}}
\caption{(Color online) Same as Fig.~\ref{fig:large} for $\xi=5$, with $A(5)\approx 3.72$, $B(5)\approx 2.92$. Agreement with the theory (\ref{eq:ABpor}) starts when $S_2 \gtrsim \xi$.}
\label{fig:x5}
\end{figure}

At larger $\xi$ one has to take into account also the subleading terms $W_p^2$ in the argument of the exponential (\ref{eq:I}); neglecting them as in Eq.(\ref{eq:Iar}) gives a rigorous upper bound on purity. One can also get a (poor) lower bound on $I(t)$ by replacing all subleading terms with the leading $W_r^2$, such that the argument of the exponential is at most $\tau^2 (L^2-r^2) W_r^2$, again resulting in a bound with a logarithmic correction. To get a better estimate we write a sum $\sum_{p=r}^{L-1} c^{(r)}_{m,p} W_p^2$ in (\ref{eq:exact}) in terms of a prefactor $x$ (that can in principle depend on $r$) as $\sum_{p=r}^{L-1} c^{(r)}_{m,p} W_p^2=:(1+x)W_r^2$. We want to account for different $x$ statistically, describing it by a probability distribution $p(x)$. One can get the exact expression for the average $\bar{x}$ by averaging over all $a_r$ arguments of the exponential function in Eq.(\ref{eq:exact}), $\frac{1}{a_r}\sum_{m,p} d_m c_{m,p}^{(r)} W_p^2=:W_r^2(1+\bar{x})$. Counting the number of times $w_p:=\sum_m d_m c_{m,r+p}^{(r)}$ one gets each $W_{r+p > r}^2$ in the $r$-th order terms, one gets, by a similar argument as used for $a_r$, that $w_p=(p+3)r2^{L-r-3}$, and as a consequence $1+\bar{x}=1+\sum_{p=1}^\infty \frac{p+3}{4} \e{-2p/\xi}=\frac{1}{16}(3+1/\tanh{(1/\xi)})^2$ irrespective of $r$ (in the TDL). The simplest improvement compared to the small-$\xi$ result would be to replace $\tau$ in Eq.(\ref{eq:Iplat}) with $\tau \sqrt{1+\bar{x}}$, i.e., just rescaling time, which can in turn be absorbed in constants $A(\xi)$ and $B(\xi)$. One can do a bit better though. Looking at a numerical distribution $p(x)$ for intermediate $\xi$, such that finite size effects for our $L$ are negligible, we find (Fig.~\ref{fig:podx}) that $p(x)=\frac{2x}{b} \e{-x^2/b}$ describes the distribution reasonably well.
Using $\bar{x}:=\int_0^\infty\!\! p(x)x {\rm d}x=\sqrt{\pi b/4}$ one can determine the needed $b$ for each $\xi$ such that $\bar{x}$ has the required exact value. Averaging over such parameter-free $p(x)$ the purity can be written as
\begin{equation}
\ave{I(t)}=\int_0^\infty{\frac{r+1}{2^{r+2}} \e{-\tau^2 (1+x)\e{-2r/\xi}} \frac{2x}{b}\e{-x^2/b}\, {\rm d}r\, {\rm d}x}.
\label{eq:Ipor}
\end{equation}
The integral over $r$ can be evaluated (it is the same as the infinite integral in Eq.~(\ref{eq:Iint}) with a rescaled $\tau \to \tau\sqrt{1+x}$), obtaining
\begin{eqnarray}
I(\tau)=&&\frac{\xi}{8}\Gamma\left(\frac{\xi}{2}\ln{2}\right)\frac{1}{\tau^{\xi \ln{2}}} \frac{1}{(1+x)^{\frac{\xi}{2}\ln{2}}} \left[1+\xi \ln{\tau}+\vphantom{\frac{\xi}{2}\ln{(1+x)}} \right. \nonumber \\
&&\left. +\frac{\xi}{2}\ln{(1+x)}-\frac{\xi}{2}\Psi\left(\frac{\xi}{2}\ln{2}\right) \right].
\end{eqnarray} 
Finally averaging this over $p(x)$ and taking the logarithm of the average in order to get $S_2$, we again obtain the familiar
\begin{eqnarray}
S_2(\tau)=\xi \ln\tau+A(\xi)-\log_2{\left[B(\xi)+\xi \ln \tau\,\right]},\nonumber \\
 A(\xi)=-\log_2{\left[\frac{\xi X_1}{8}\Gamma\left(\frac{\xi}{2}\ln{2}\right)\right]},\nonumber \\
 B(\xi)=1-\frac{\xi}{2} \Psi\left(\frac{\xi}{2}\ln{2}\right)+\frac{\xi X_2}{2X_1},\nonumber \\
X_1:=\int_0^\infty \frac{1}{(1+x)^{\frac{\xi}{2}\ln{2}}}\frac{2x}{b} \e{-x^2/b} {\rm d}b, \nonumber\\
X_2:=\int_0^\infty \frac{\ln{(1+x)}}{(1+x)^{\frac{\xi}{2}\ln{2}}}\frac{2x}{b} \e{-x^2/b} {\rm d}b,
\label{eq:ABpor}
\end{eqnarray}
where $b$ is determined from $\xi$ by $\sqrt{\pi b/4}=\frac{1}{16}(3+1/\tanh{(1/\xi)})^2-1$. The form that we get for $S_2$ is the same as in Eq.(\ref{eq:Iplat} though with a modified $A(\xi)$ and $B(\xi)$, shown also in Fig.~\ref{fig:AB2}.
In Figs.~\ref{fig:large} and~\ref{fig:x5} we show comparison between the exact $S_2$ and our approximate result (\ref{eq:ABpor}), finding good agreement for all $\xi$ that we checked (small and large). Note that for larger $\xi$ theoretical form Eq.~(\ref{eq:ABpor}) describes $S_2$ well for not too short times such that $S_2 \gtrsim \xi$. In order so clearly see the logarithmic correction we also show a logarithmic derivative of $S_2$ (\ref{eq:Iplat}) that behaves as ${\rm d}S_2/{\rm d}(\ln t)=\xi-\xi/[(B+\xi\ln{\tau})\ln{2}]$, that is, it approaches the asymptotic $\xi$ with finite-time correction of order $\sim 1/\ln{\tau}$. Due to boundary effects we could not check even larger localization lengths $\xi  \gg 5$, where, if the distribution $p(x)$ would change, the values of $A$ and $B$ could be modified.

We have throughout focused on a particular initial state and a Gaussian distribution of couplings. We show in Appendix~\ref{app1} that a different generic initial product state, different distribution of coupling constants, as well as using von Neumann entropy instead of $S_2$, leads to essentially the same results as our exact calculation for a particular initial state and $S_2$.

\section{The 3-body random model}

A natural question is if any of the results obtained for the 2-body random model, in particular the logarithmic correction, could be modified by higher $r$-body diagonal interactions (\ref{eq:H}), which, though being less important (for an MBL phase $J^{(r)}$ should decay exponentially in $r$), could bring some fundamentally different behavior. As we explained, because the correction essentially comes from the simple geometrical bonds counting, this is unlikely. In the following we present exact results demonstrating that.

To this end we consider a pure 3-body random model, where only $J^{(3)}_{k,l,m}$ are nonzero i.i.d. Gaussian numbers with zero mean and the variance 
\begin{equation}
\ave{(J^{(3)}_{k,l,m})^2}=J^2 W_{r}^2=J^2 {\rm e}^{-2(m-k-1)/\xi}, \quad \hbox{$k<l<m$},
\end{equation}
where, as before, $r-1=m-k-1$. Taking the initial state that is a uniform mixture of all basis states, $c_{i\alpha}=1/\sqrt{N}$, one gets
\begin{eqnarray}
I(t)=&&\frac{1}{N^2}\!\!\!\! \sum_{\substack{i,j\in A \\\alpha,\beta \in B}}\!\!\!\! {\rm exp}\!\!\left(\!\!-\ii t\! \left[\! \sum_{\substack{l,m\in B\\k \in A}}\!\!\! (i_k-j_k)J^{(3)}_{k,lm} (\alpha_l \alpha_m-\beta_l \beta_m)+\nonumber \right.\right.\\
&&\left. \left.+ \sum_{\substack{k,l \in A\\ m\in B}} (i_k i_l-j_k j_l)J^{(3)}_{kl,m}(\alpha_m-\beta_m) \right]\right),
\label{eq:z3}
\end{eqnarray}
where $k,l,m$ are site indices while $i,j,\alpha,\beta$ are multiindices labeling the basis. Compared to the 2-body model the saturation value has an exponentially small correction and is for an equal bipartite cut $L_{\rm A}=L/2$ equal to $\overline{I(t)}=2/2^{L/2}$. Averaging over Gaussian distribution of couplings we get a sum of Gaussian functions, like in (\ref{eq:exact}), though with a more complicated combinatorics of $c$'s and $d$'s. As an example, for $L=2L_{\rm A}=4$ sites averaging (\ref{eq:z3}) over Gaussian $J^{(3)}$ gives the exact expression
\begin{equation}
I(t)=\frac{1}{16}\left[8+ \e{-2\tau^2 W_2^2}+\e{-2\tau^2 W_3^2}+6\e{-\tau^2(W_2^2+W_3^2)} \right].
\label{eq:3L4}
\end{equation}
Complexity of the exact expression of course again grows with $L$ so we shall focus on small $\xi$ case where one can again neglect the subleading terms $W_{l>r}^2$. 

Counting the number of terms with the leading order $r$ one gets $a_r=(r2^r+6)2^{L-2r-1}=r2^{L-r-1}(1+\frac{6}{r2^r})$, for $r=2,\ldots L/2$, and $a_r=(L-r)2^{L-r-1}$ otherwise (note that for the 3-body model the smallest distance is $r=2$). Therefore, asymptotically for large $r$ the form of $a_r$ is the same as for the 2-body random model. What is different though is that the leading prefactor $c^{(r)}_{m,r}$ is not $1$ like in Eq.~(\ref{eq:exact}). Some of the $a_r$ terms have a prefactor $c^{(r)}_{m,r}=r-1$, some $c^{(r)}_{m,r}=2(r-1)$. The fraction of those with $c^{(r)}_{m,r}=r-1$ is equal to $\frac{2+r2^{r-1}}{3+r2^{r-1}}$ which goes to $1$ for large $r$ (e.g., for $r=2$ it is $\frac{6}{7}$). For instance, in the above $L=4$ case (\ref{eq:3L4}) we have $a_2=7$ and $a_3=1$, out of all $a_2$ terms $\frac{6}{7}\cdot 7=6$ have a prefactor in the exponential $c=1$, while $1$ has $c=2$. We shall therefore neglect terms with $c^{(r)}_{m,r}=2(r-1)$, writing the average purity
\begin{equation}
\ave{I(t)}\approx \sum_{r=2}^{L-1} \frac{a_r}{2^{L}} \e{-(r-1)\tau^2W_r^2} \asymp \sum_{r=2}^\infty \frac{r}{2^{r+1}} \e{-(r-1)\tau^2 W_r^2}.
\label{eq:I3}
\end{equation}
Compared to the 2-body model the only difference is an additional factor $r-1$ in front of $\tau^2$. 
\begin{figure}[tb!]
\centerline{\includegraphics[width=3in]{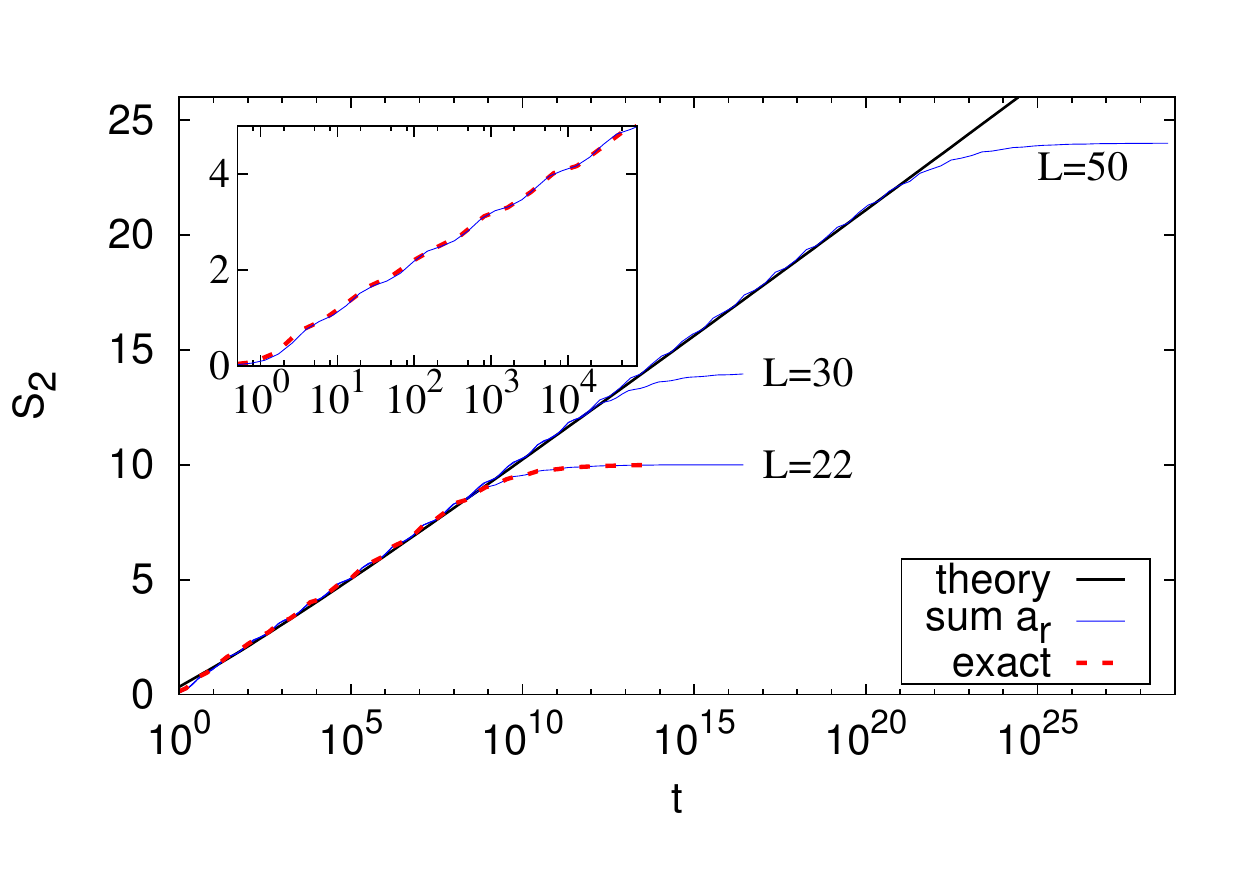}}
\caption{(Color online) Random 3-body dephasing model for $\xi=0.5$. Exact entanglement $S_2$ (red dashed curves) almost overlaps with the approximation (\ref{eq:I3}) and with theory (\ref{eq:3Q}).}
\label{fig:3Q}
\end{figure}
Calculating the time when $I(t)$ hits the middle between two consecutive plateaus, similarly as for the 2-body model, one gets $r \approx \xi \ln{\tau_r}+1+\frac{\xi}{2}\ln{(\xi \ln\tau_r)}$, resulting in
\begin{eqnarray}
S_2(\tau)\approx&&\,y+\frac{3}{2}-\log_2{\left(y+\frac{5}{2} \right)},\nonumber \\
&& y:=\xi\ln\left[\tau (\xi\ln{\tau})^{p/2} \right],
\label{eq:3Q}
\end{eqnarray}
where $p=1$ for the 3-body model, coming from a $c_{m,r}^{(r)}=(r-1)^{p}$, while the 2-body small-$\xi$ result (\ref{eq:Iplat}) is obtained for $p=0$. In Fig.~\ref{fig:3Q} we can see nice agreement of Eq.(\ref{eq:I3}) and (\ref{eq:3Q}) with the exact numerical calculation of $S_2$. In terms of the scaling variable $y$ the form is the same as for the 2-body model, where the scaling variable was $y_{\rm 2-b}:=\xi \ln{\tau}$. Expanding the logarithm we have $y=\xi \ln{\tau}+\frac{\xi}{2}\ln{(\xi \ln{\tau})}$, and therefore, compared to the 2-body result (\ref{eq:Iplat}), there is an additional logarithmic correction proportional to $\xi$. The same holds for any other finite $p$, and even after averaging over different $p$ one would get at most $S_2 = u+A(\xi)+C(\xi)\ln{u}-\log_2{[u+B(\xi)+C(\xi)\ln{u}]}$, where $u:=\xi \ln{\tau}$. Therefore, we conjecture that any $r$-particle interaction (with finite $r$) can not fundamentally alter logarithmic corrections that we have found. One difference though worth mentioning is that for large $\xi$ (or $p$) the positive prefactor $C(\xi)$ of the first correction can be larger than $1/\ln{2}$ (a negative prefactor of the 2nd correction) and so in total the corrections can be positive instead of negative -- entanglement growth is slightly faster than logarithmic. In the 2-body model it was always slightly slower.

\section{Many-body localization}

So-far we have calculated the evolution of entanglement in the l-bit basis, how about the original physical basis of an MBL system that can be transformed to the l-bit form? For that one has to apply a basis rotation at $t=0$, and at final $t$. Because it is a quasi-local unitary that transforms between the two bases, i.e., a finite-depth circuit~\cite{Bauer13}, it can modify the entanglement $S_2$ only by a constant term that is proportional to the circuit depth/localization length (and is independent of time). For long times when $S_2$ is large this can not modify neither the leading log term, nor the subleading correction because they both grow with time. In a realistic MBL system, unless the localization length is very small, many different $r$-body terms in $H$ will contribute. While, as we argued, this will not change the form of the subleading correction, it will influence constants $A(\xi),B(\xi),C(\xi)$, as well as likely wash-out sharp plateaus in the growth that we observed for small localization length.

Checking for possible sub-leading corrections that are present in the l-bit model in a given concrete $H$ that is believed to display MBL is an interesting problem that would give information on whether the l-bit picture is indeed an exact one. Such a study however goes beyond the scope of the present paper. To unambiguously identify a sub-leading term one will need large systems as well as large times (see e.g. Fig.~\ref{fig:small}). Simply doing exact diagonalization on say $16$ spins in a double precision floating point arithmetic will likely not suffice.

\section{Spectrum evolution}
\begin{figure}[t!]
\centerline{\includegraphics[width=3.2in]{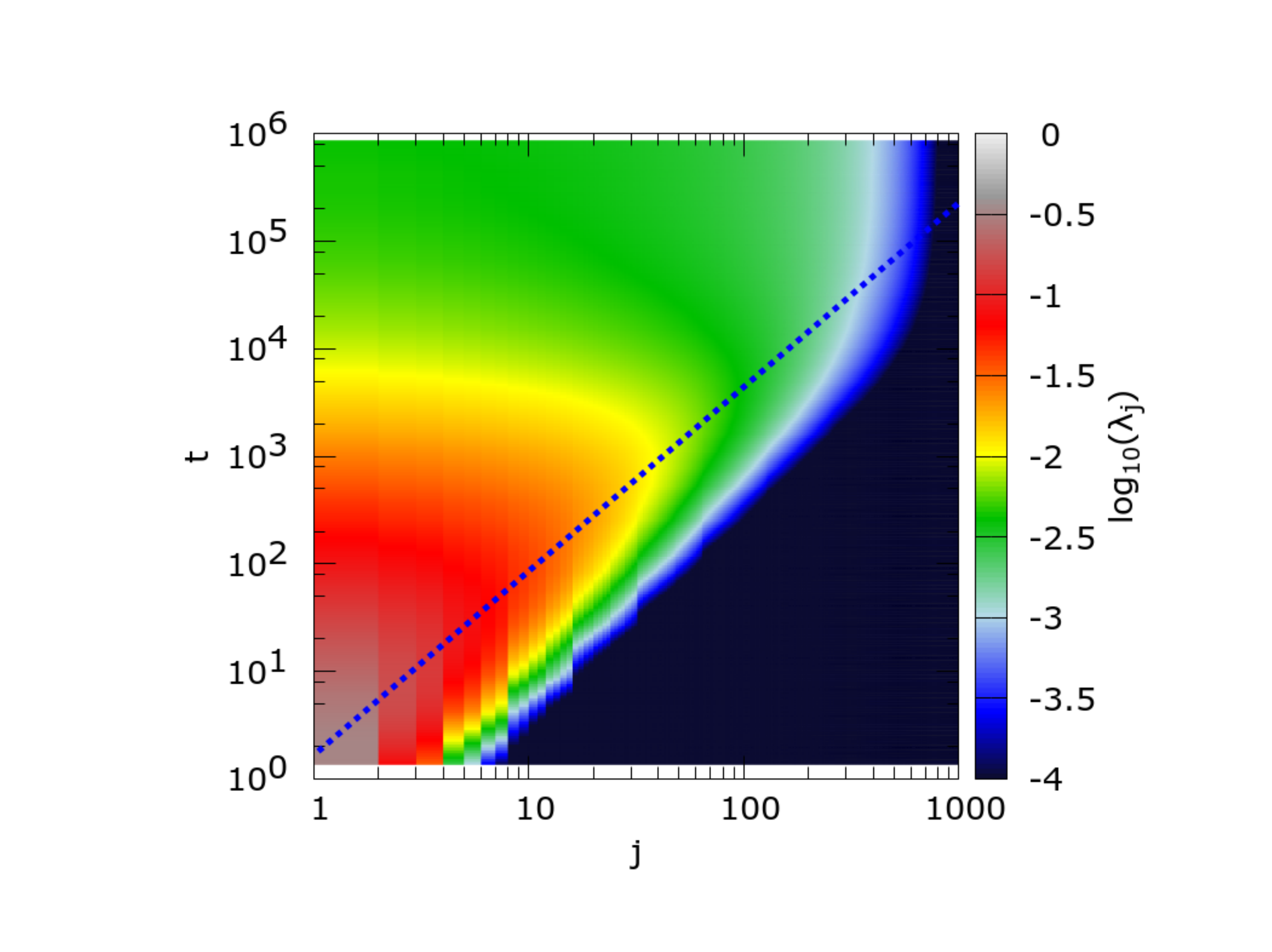}}
\caption{(Color online) Time evolution of the reduced density matrix eigenvalues $\lambda_j$ for a half-cut and the 2-body dephasing model with $\xi=1$ and $L=20$. Dashed blue line is (see also Fig.~\ref{fig:lam2}) $t_{\rm max}\approx 1.8 j^{1.7}$, giving the location of the maxima of $\lambda_j(t)$.}
\label{fig:lam1}
\end{figure}
Full information about entanglement properties of a given pure state is contained in the spectrum $\lambda_j$ of $\rho_{\rm A}(t)$. While $S_2$, being a scalar quantity that depends on eigenvalues $\lambda_j$, subsumes overall evolution of entanglement we here consider also dependence of each individual $\lambda_j(t)$, $j=0,\ldots N_{\rm A}-1$, ordered nonincreasingly, $\lambda_j\ge \lambda_{j+1}$. We focus here just on the leading order behavior as it gives some interesting effects that have not been studied before.

\begin{figure}[tbh!]
\centerline{\includegraphics[width=2.7in]{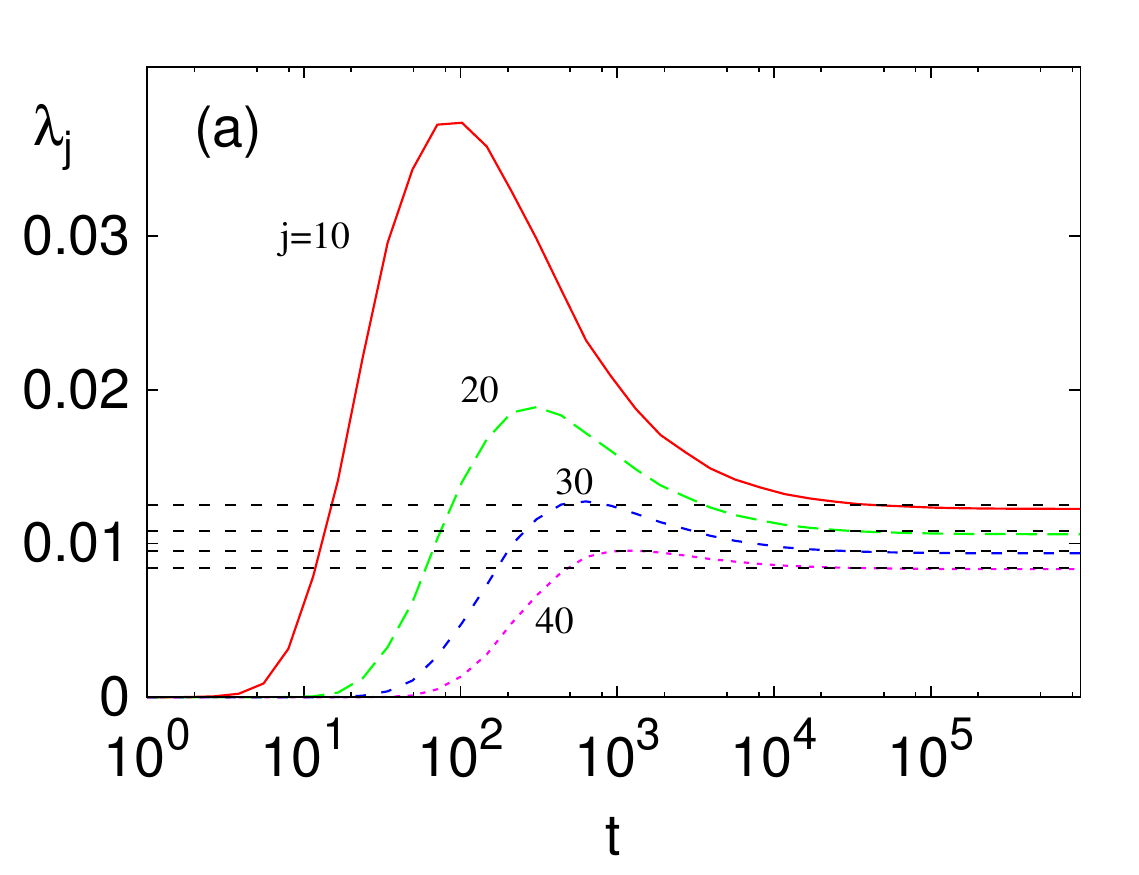}}
\centerline{\includegraphics[width=2.73in]{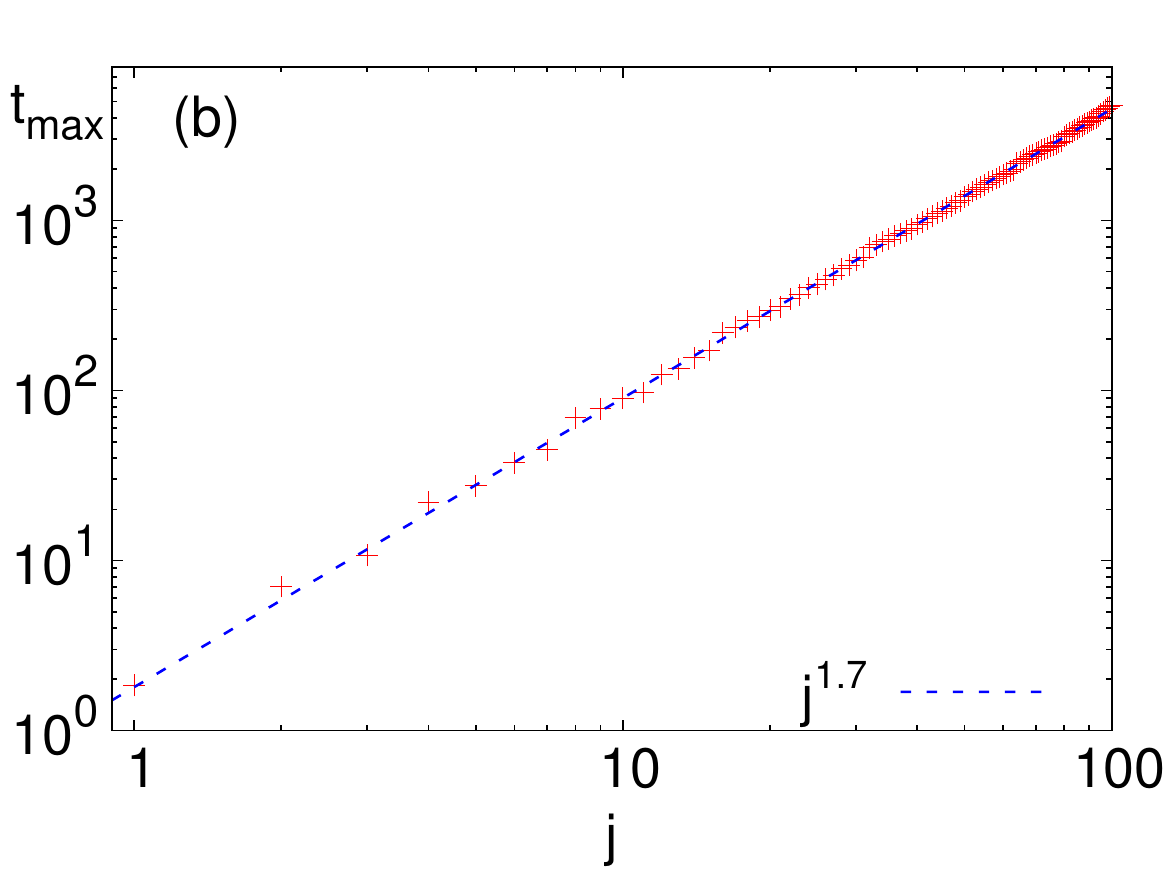}}
\caption{(Color online) Same as in Fig.~\ref{fig:lam1}. (a) Few selected $\lambda_j(t)$ for $L=16$. Horizontal lines are theoretical values~\cite{MZ07} (see text) for random states. (b) Time of the maxima of $\lambda_j(t)$ for different $j$ (time $t_{\rm max}$ is independent of $L$; we show $L=20$, the same data as in Fig.~\ref{fig:lam1}).}
\label{fig:lam2}
\end{figure}

Results of numerical simulation for the 2-body random model are shown in Figs.~\ref{fig:lam1} and~\ref{fig:lam2}. Looking at the time dependence of $\lambda_j$ (Fig.~\ref{fig:lam2}a) we can see that they have a nonmonotonic dependence (except the largest one $\lambda_0$, data not shown), with a single maximum achieved at $t_{\rm max}$, while at large time they approach $\lambda_j$ for random states, given implicitly~\cite{MZ07} by $2^{L/2}\lambda_j=4\cos^2{\varphi_j}$, where $\frac{\pi}{2}\frac{j+0.5}{2^{L/2}}=\varphi_j-\frac{1}{2}\sin{(2\varphi_j)}$. These long-time values are the same as for the evolution by completely random diagonal matrices~\cite{Karol}. Time $t_{\rm max}$ strongly depends on $j$, i.e., the larger eigenvalues ``turn on'' the fastest (which is different than in generic evolution modeled by a random matrix~\cite{Vinayak}), the dependence being $t_{\rm max}(j) \approx f(\xi)\,j^{1.7/\xi}$. The value at the maximum $\lambda_j(t_{\rm max}(j))$ is on the other hand $\sim 1/j$ and is independent of $\xi$ and $L$ (this is in line with a generic volume-law states reached at that late time). Therefore, quantum correlations and the entanglement rank increases gradually from larger $\lambda_j$ to smaller, in line with exponentially decaying couplings. We note that the power $1.7/\xi$ in $t_{\rm max} \sim j^{1.7/\xi}$ is smaller for larger $\xi$. It can be explained simply from the scaling of distance $r-1 \sim \xi \ln{t}$, which relates to $j \sim 2^r$ eigenvalues being nonzero at that time. This results in a scaling $t \sim j^{1/(\xi \ln 2)}$. That the prefactor $1/\ln{2} \approx 1.44$ is not exactly $1.7$ is likely because of certain arbitrariness in choosing the precise time that we look at. After time $t_{\rm max}$ the eigenvalues gradually relax to their random-state asymptotic values. The localization (decay) length $\xi$ can therefore also be inferred from the location of the maxima of individual eigenvalues.

We end by noting that the entanglement spectrum of eigenstates has proved to be useful for understanding MBL before~\cite{Serbyn16}, as well as the Schmidt gap~\cite{Bose} and the distribution of entanglement~\cite{Sing16}. Further details of the evolution of $\lambda_j(t)$ need to be studied in the future.

\section{Conclusion}

We have calculated the exact form of entanglement evolution (within well controlled approximations) for a diagonal l-bit model with random exponentially decaying couplings that are supposed to describes one-dimensional many-body localized phase, showing that the ``established'' logarithmic growth is in fact not exact. Solving a 2-body and a 3-body model we find that there is an additional log-log correction that comes essentially from the linear growth with length of the number of couplings connecting two bipartitions. As a consequence, the entanglement growth is slightly slower at shorter than at longer times. The result is robust and should be present in any exponentially localized many-body phase describable by the l-bit (LIOM) model. 

This constitutes one of few analytical results for many-body localized systems. As such it should be valuable as a benchmark property of a widely accepted characterization of localization through correlations spreading. The techniques used can be generalized to more than one dimension. Finding a so-far unknown contribution shows the importance of pursuing exact calculations also for other quantities, thereby providing a more solid footing for the many-body localization against which results of numerical studies can be compared.

\section*{Acknowledgements}
This work is supported by Grants No. J1-7279 and P1-0044 from the Slovenian Research Agency.

\appendix

\section{Numerical checks}
\label{app1}

\begin{figure}[thb!]
\centerline{\includegraphics[width=3in]{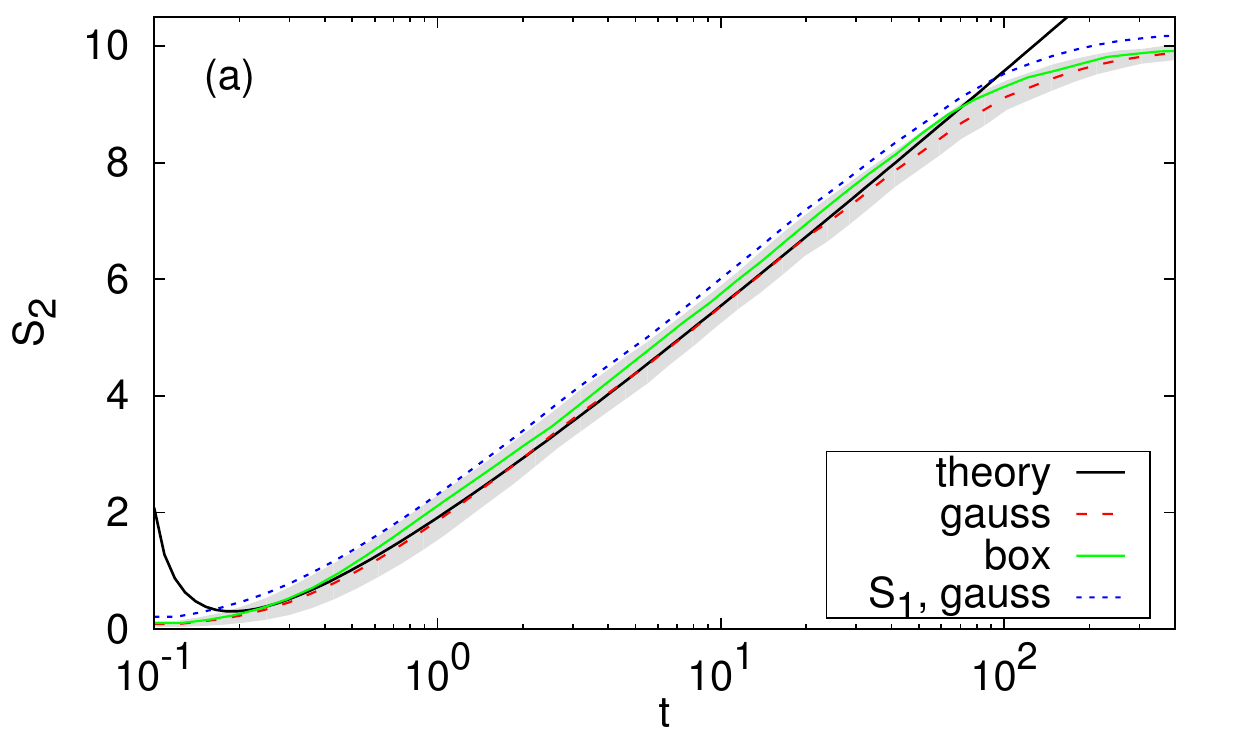}}
\centerline{\includegraphics[width=3in]{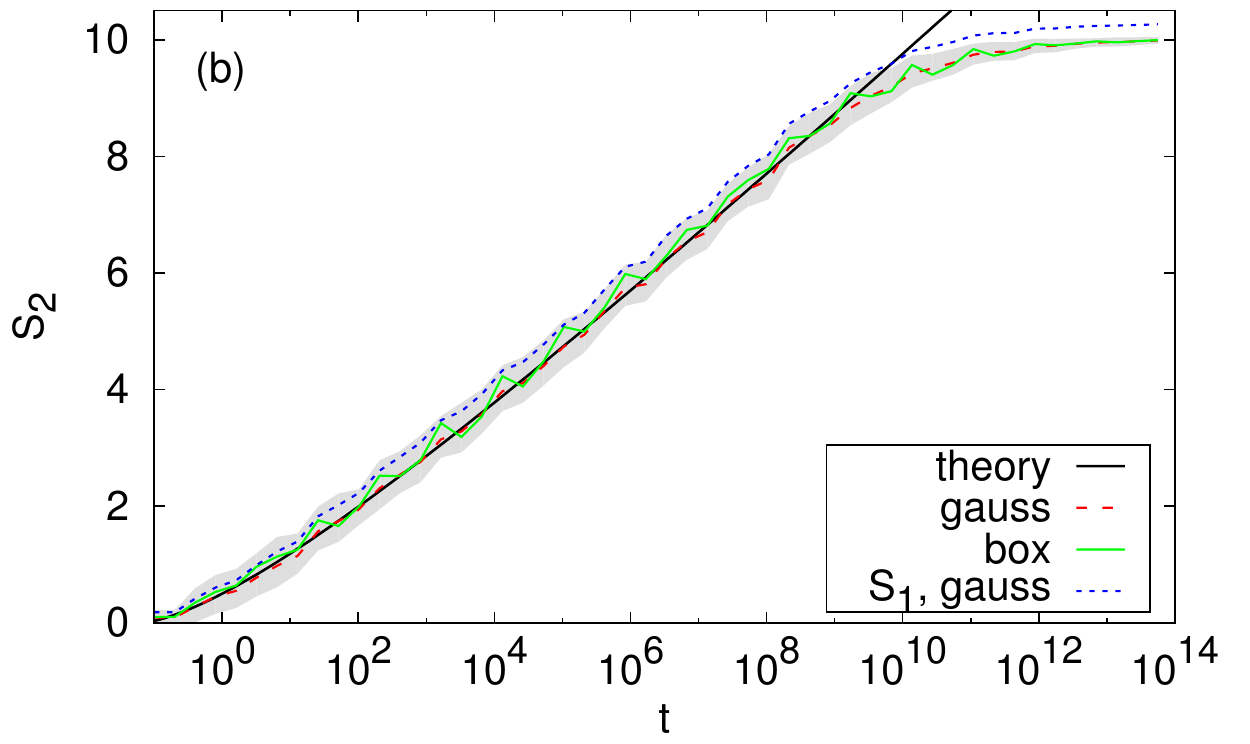}}
\caption{(Color online) Comparison of $S_2$ for a Gaussian distribution of couplings and a box distribution (with the same variance). (a) is for $\xi=2$, (b) for $\xi=0.5$, both for $L=22$. Theory is (\ref{eq:ABpor}), gray shading denotes standard deviation of $S_2$ for a Gaussian case, while blue dotted curves show von Neumann entropy $S_1$.}
\label{fig:box}
\end{figure}
Here we verify that the physics of entanglement growth in the 2-body random model is essentially the same also for other distributions of $J^{(2)}$, von Neumann entropy, and initial states that do not have simple $c_{j\alpha}=1/\sqrt{N}$.

\begin{figure}[t!]
\centerline{\includegraphics[width=3in]{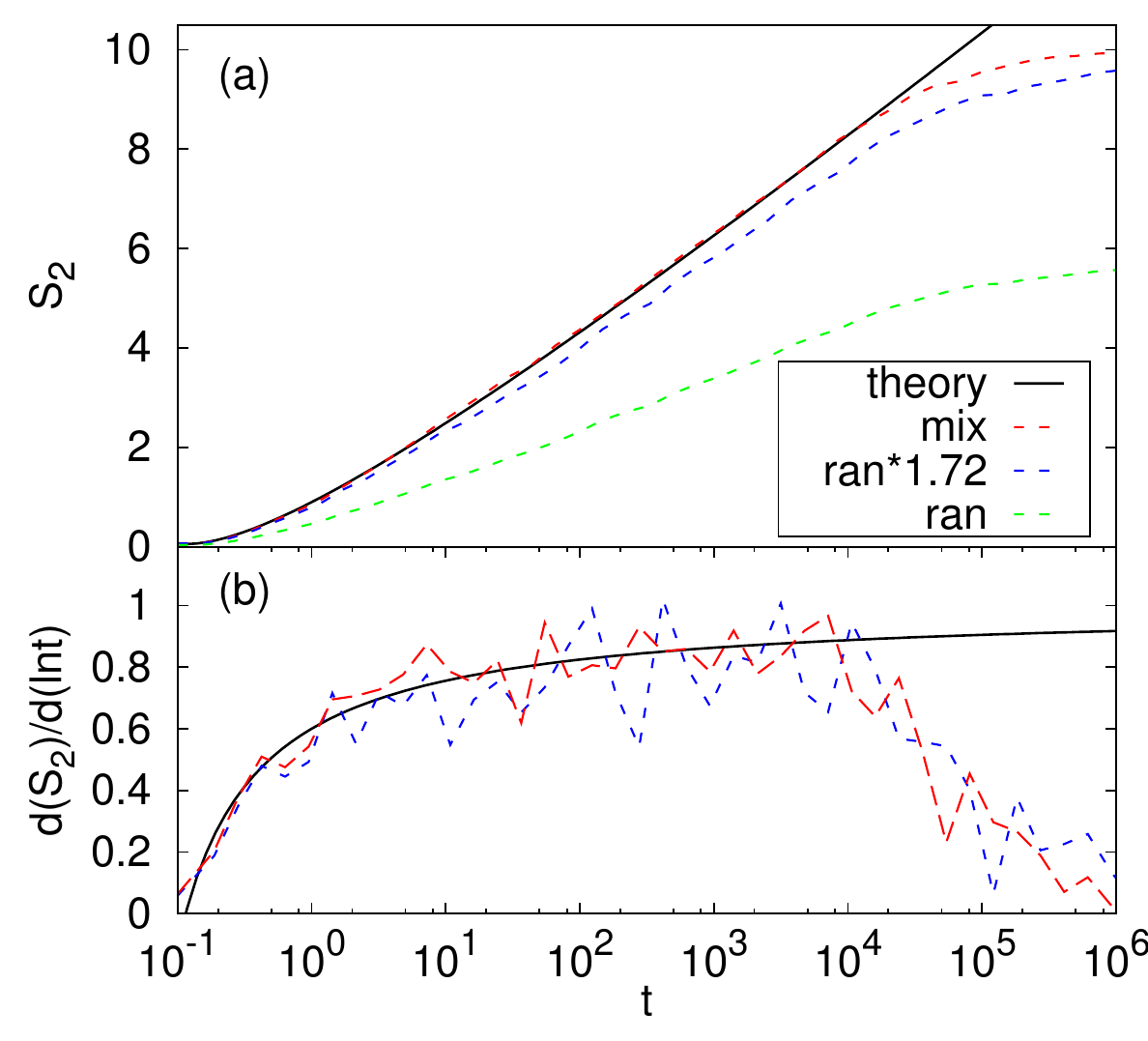}}
\caption{(Color online) Comparison of $S_2$ (in (a)) and its derivative (in (b)) for a uniform mixture initial state, $c_{j\alpha}=1/\sqrt{N}$ (red dashed curve, labeled ``mix''), and random product states on the Bloch sphere (labeled ``ran'', blue and green curve), all for $L=22$ and $\xi=1$. Blue dashed curve is data for the green one multiplied by $1.72 \approx (1\cdot11-1)/(0.62\cdot 11-1)$.}
\label{fig:ran}
\end{figure}
In Fig.~\ref{fig:box} we compare numerically calculated $S_2$ for a Gaussian distribution of couplings that was discussed in the main text, and for which analytics is the simplest, and a box distribution. We can see that the behavior is essentially the same. For a non-Gaussian distribution in Eq.(\ref{eq:exact}) one would for instance instead of a sum of Gaussian functions have a sum of Fourier transformations of the distribution. For the box distribution and small $\xi$ (e.g., frame (b) in Fig.~\ref{fig:box}) one can see a non-monotonic increase of $S_2$ that is due to the oscillating nature of the Fourier transformation of the box distribution -- any distribution with a finite support will exhibit such diffrative oscillations. We can also see that, as anticipated, the relative variance $\sigma(S_2)/S_2$ goes to zero for large times. Volume-law states that appear at late times are self-averaging and one can replace the average of the logarithm with a logarithm of the average, as we have done throughout. Last point to note in Fig.~\ref{fig:box} is that the von Neumann entropy, denoted by $S_1$, behaves similarly as $S_2$.

We also check different initial states. We argued that the choice $c_{j\alpha}=1/\sqrt{N}$ was mostly for analytical convenience and that other generic product initial state choices should result in the same entropy growth. From a finite-size effects point of view it is best to choose product initial states that have zero expectation of $\sz{j}$, $z=0$, as this results in the largest saturation value $S_2(t \to \infty)=c\frac{L}{2}-1$, where $c=\log_2\frac{2}{1+z^2}$. As long as the initial state is a product state with all single-site orientations being in the $x-y$ plane, the results presented are still exact after averaging over $i.i.d.$ distribution of orientations within the $x-y$ plane. For initial states that have random uniform orientation on the Bloch sphere the saturation value will be smaller; it can be estimated by averaging $c$ over the Bloch sphere, giving $\int_0^1 \log_2{\frac{2}{1+z^2}} {\rm d}z=\frac{4-\pi}{\ln{4}}\approx 0.62$. In Fig.~\ref{fig:ran} we compare data for a random product initial state and a state with $c_{j\alpha}=1/\sqrt{N}$, seeing that after we rescale the random-state data by the theoretical factor accounting for different asymptotic saturation values, the two behave essentially the same, including the logarithmic correction (frame (b)).


\begin{thebibliography}{}
\bibitem{foot1} We remark that the logarithmic growth of entanglement can not be used as a sole indicator of MBL. There are systems that display logarithmic complexity growth but are not MBL, e.g., non-interacting systems~\cite{Rigol17,Hetterich17} with bond disorder~\cite{Montangero06} or long-range hopping~\cite{Roy17}, or even clean systems~\cite{Yao16,our}; there are also systems without an explicit disorder that have slow double-logarithmic growth~\cite{Anto17}.
\end{thebibliography}
\end{document}